\documentclass{ws-ijmpa}

\begin{document}

\markboth{B. Mota, M. J. Rebou\c{c}as and R. Tavakol} {Local Shape
of the Universe}

\catchline{}{}{}{}{}

\title{THE LOCAL SHAPE OF THE UNIVERSE \\ IN
              THE INFLATIONARY LIMIT}

\author{\footnotesize B. MOTA, \ M. J.
REBOU\c{C}AS}

\address{Centro Brasileiro de Pesquisas F\'{\i}sicas
\\Rua Dr.\ Xavier Sigaud 150, 22290-180 Rio de Janeiro -- RJ,
Brazil\\brunom@cbpf.br,\ reboucas@cbpf.br}

\author{\footnotesize R. TAVAKOL}

\address{Astronomy Unit, School of Mathematical Sciences,
Queen Mary, University of London
\\ Mile End Road, London E1 4NS, UK\\r.tavakol@qmwul.ac.uk}

\maketitle

\pub{Received (Day Month Year)}{Revised (Day Month Year)}

\begin{abstract}

Recent  high precision data by WMAP and SDSS have provided strong
evidence to suggest that the universe is nearly flat. They are
also making it possible to probe the topology of the universe.
Motivated by these results, we have recently studied the
consequences of taking the inflationary limit, i.e. $|\Omega_{0} -
1| \ll 1$. We have shown that in this limit a generic detectable
spherical or hyperbolic topology is locally indistinguishable from
either $\mathbb{R}^{2} \times\mathbb{S}^{1}$ or 
$\mathbb{R}\times\mathbb{T}^{2}$, irrespective of its global
shape. Here we briefly present these results and further discuss
their observational implications.

\end{abstract}


\section{Introduction}

Among the most fundamental questions in cosmology are those
relating to the natures of the geometry and topology of the
universe. The geometry can be obtained from the matter-energy
content of the universe, through Einstein's field equations. In
this general relativistic context, the universe seems to be well
described by a locally homogeneous and isotropic
Friedmann-Lema\^{\i}tre-Robertson-Walker (FLRW) metric, with a
constant spatial curvature with sign $k=0$, $\pm1$.

General relativity is however a local metrical theory, and as such
does not determine the topology of spacetime. In general, the
spatial section of a FLRW spacetime can be a multiply connected
manifold (which we assume compact and orientable)
$M=\widetilde{M}/\Gamma$, where the covering space $\widetilde{M}$
is either $\widetilde{M}=\mathbb{E}^{3}$, $\mathbb{S}^{3}$ or
$\mathbb{H}^{3}$, depending upon $k$; and $\Gamma$ is a discrete
and  fixed-point free group of isometries of $\widetilde{M}$.
Thus, unless specified by a fundamental theory, the topology of
the universe is expected to be inferred from observations.\cite{SomeTopRefs} 
An immediate consequence of such multiple-connectedness, is that
there will be multiple images of each radiating source, one for
each isometry in $\Gamma$. Therefore, searching for a non-trivial
topology amounts to a search for multiple images or pattern
repetitions. Due to the existence of a cosmological horizon,
however, only the subset of images within the observable universe
(i.e. a sphere of radius $\chi_{obs}$, where $\chi_{obs}$ is the
redshift--distance relation evaluated at $z=z_{max}$ of the survey
used, centered at the observer's position) is
detectable.\cite{grt}

Most studies of cosmic topology so far have concentrated on
particular manifolds. Given that there are an infinite number of
possible topologies for each curvature sign, a more reasonable
search strategy is to ask what is the set of all detectable
topologies. In general, each candidate manifold has a distinct
isometry group $\Gamma$. Since not all isometries give rise to
detectable multiple images, 
it is reasonable to restrict our focus only to the subgroup
generated by \textit{detectable} isometries, which we refer to as
the \textit{local shape of the universe}. As we shall show, this
greatly simplifies the search for image repetitions, while making
it possible to obtain some very general results. In the following
we briefly outline our recent results for the spherical
spaces,\cite{our} which show that in an appropriate limit typical
manifolds have approximately the same local shape, and further
discuss their observational implications.

\section{Main Results}

Recent high precision data by WMAP\cite{WMAP} are making it
possible to comprehensively probe the topology of the universe for
the first time, using pattern repetition. In addition they have
provided strong evidence suggesting that the universe is nearly
flat, with $\Omega_{0}\simeq1$. According to the most recent
estimates of the density parameters, $\Omega_{0}\in[0.99,1.03]$
with a $2\sigma$ confidence,\cite{SDSS} which still allows the
3-space to be spherical, hyperbolic or flat. Furthermore, in a
non-flat inflationary universe one would typically expect
$|1-\Omega_{0}|\ll1$, implying  $\chi_{obs}\ll1$ in units of the
curvature radius. In addition, non-flat manifolds are rigid, in
the sense that all metrical quantities, such as the length of
closed geodesics, are fixed in units of the curvature radius.
Hence, the inflationary limit imposes important constraints on the
set of detectable isometries for non-flat manifolds, and hence on
their local shape.

Our aim here is to take a general point of view and ask what are
the set of topologies that would be detectable in very nearly flat
universes. Thus to proceed we shall, in addition to assuming
$|\Omega_{0}-1|\ll1$, make two further physically motivated
assumptions: (i) the observer is at a position $x$ where the
topology is detectable, i.e. $r_{inj}(x)<\chi_{obs}$, where
$r_{inj}(x)$ is the injectivity radius at $x$ defined as half the
length of the smallest closed geodesic passing through $x$ (see
Ref. \refcite{our} for details), and (ii) the topology is not
excludable, i.e. it does not produce too many images so as to be
ruled out by present observations. Thus, our main physical
assumption can be summarized as
\begin{equation}
r_{inj}(x)\lesssim\chi_{obs}\ll1\;.\label{hip}%
\end{equation}
These assumptions severely restricts the set of detectable
non-flat manifolds. Thus in the case of spherical manifolds, only
lens spaces (with $r_{inj}=\frac{\pi}{p}$) and binary dihedrical
spaces (with $r_{inj}=\frac{\pi}{m}$) of sufficiently high order
of $p$ or $m$ are detectable. In the hyperbolic case, the only
detectable manifolds are the so-called nearly cusped manifolds,
which are sufficiently similar to the cusped manifolds (cusped
manifolds are not compact, and possess regions with arbitrarily
small $r_{inj}(x)$.)

In a recent study\cite{our} we considered both classes of
manifolds and showed that a generic detectable spherical or
hyperbolic manifold is locally indistinguishable from either a
cylindrical ($\mathbb{R}^{2} \times\mathbb{S}^{1}$) or toroidal
($\mathbb{R}\times\mathbb{T}^{2}$) manifold, irrespective of its
global shape. Here, we shall briefly review our results, and
further discuss their observational implications.

Briefly the key arguments are the following. At any given point
$x$, an isometry $\gamma$ will generate a closed geodesic with
length $\ell=d(x,\gamma x)$. But given the limit, (\ref{hip}) we
are only interested in geodesics short enough so that both the
sources and their images lie within the cosmological horizon. The
question then is what do these detectable closed geodesic look
like? In particular how much do they deviate from the Clifford
translations (i.e. isometries with constant distance function $d$)
within the observable universe? We have found that for a lens
space $L(p,q)$ with generator $g$, the following results hold:

\noindent i) The manifold $L(p,q)$ is equivalent to the manifold
$L(p_{N-1},q)$, where $p_{N-1}$ is the second to last convergent
in the continued fractions expansion of $p/q$.

\noindent ii) The shortest geodesic at any point is generated by
$g^{p_{j}}$, for some convergent $p_{j}$.

\noindent iii) $r_{inj}(x)\leq\frac{\pi}{\sqrt{p}}$. Thus by
choosing $p$ to be sufficiently large, one can at each point of a
lens space intersects a closed geodesic which is small in units of
the curvature radius. We employ this bound to divide the $(q, p)$
plane into detectable, undetectable and observationally excluded
regions, as shown in  (Fig. 1a). Thus any systematic search for a
lens space-shaped universe needs only to concentrate on manifolds
that lie in the white central region of the diagram.

\noindent iv) Consider an observable sphere of radius
$\chi_{obs}$, centered at an observer's position $x_0$ (Fig. 1b).
The isometry $g$ identifies points in two faces of the fundamental
polyhedron (dashed lines, shown edgewise for simplicity). Let
$d_{0}$ be the distance $d(x_0,g x_0)$ and let $d_{\max}$ and
$d_{\min}$ denote the maximum and minimum values of $d(x,g x)$ for
a point $x$ within the detectable sphere. We then find
\begin{equation}
\frac{\Delta d}{d_{0}}\leq2\,
\frac{1}{|z_{2}||z_{1}|}\,\chi_{obs}\ll1\;,
\end{equation}
where  $|z_{1}|$ and $|z_{2}|$ are position dependent, obeying
$|z_{1}|^2+|z_{1}|^2=1$. A similar result holds for the cusp-like
parts of nearly cusped hyperbolic manifolds, namely
\begin{equation}
\frac{\Delta d}{d_{0}}\leq2\,\chi_{obs}\ll1\;.
\end{equation}
The above inequalities show clearly that subject to condition
(\ref{hip}), the detectable isometries of lens spaces (as well as
cusp-like hyperbolic manifolds) are very close to Clifford
translations for most observers. We also note that all isometries
in the isometry groups of the binary dihedral spaces are Clifford
translations, and therefore trivially obey the inequality above.
\begin{figure}[!htb]
\centerline{\def\epsfsize#1#2{0.65
#1}\epsffile{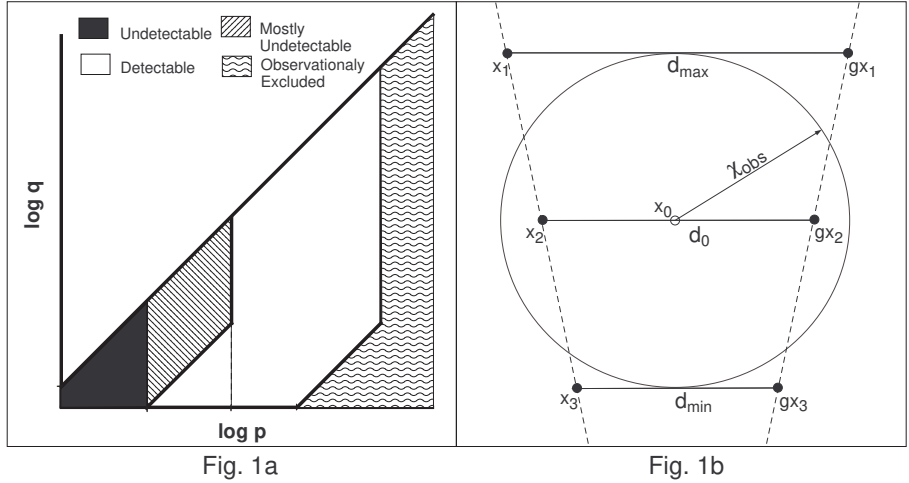}}
\caption{\label{fig=0.02} Detectability of lens spaces $L(p,q)$ in
terms of the parameters $p$ and $q$. A pattern-repetition search
should focus on manifolds on the white region (1a); and maximum
and minimum lengths of closed geodesics generated by an isometry
inside the observable universe of radius $\chi_{obs}$ (1b)}
\end{figure}
These results have important consequences for developing search
strategies for cosmic topology. They show that for a typical
observer in a very nearly flat universe, the 'detectable part' of
the topology would be indistinguishable from either
$\mathbb{R}^{2} \times\mathbb{S}^{1}$ or $\mathbb{R}
\times\mathbb{T}^{2}$ manifold. Note that the detectable
isometries cannot have more than 2 independent generators because
the fundamental domain has a far greater volume than the
observable universe. We also emphasize that no matter how flat the
universe turns out to be, there are always an infinite number of
candidate manifolds, both spherical and hyperbolic, that have
detectable isometries. Furthermore, in the case of (detectable)
spherical manifolds, we have also proven that through any point in
the manifold there is a closed geodesic of length much smaller
than the curvature radius, and therefore the fraction of the total
manifold's volume where there are detectable isometries is indeed
large.

So far our discussion has assumed the inflationary limit,
$|\Omega_0-1|\ll 1$. However, the resolutions required to test
such limits are not expected to be attainable in near future.
Currently the best fit value for $\Omega_0$ is $1.02\pm0.02$ to
$1\sigma$.\cite{WMAP} These bound are also compatible with other
topologies not considered here, such as the recently proposed
Poincar\'{e} dodecahedral space.\cite{poinc} It is therefore
important to ask what the present results can tell us for less
restrictive bounds. Results (i)-(iii) hold regardless of the value
of $\Omega_0$. The bounds in (iv) however rely on approximations
that remain valid only if $|\Omega_0-1|\lesssim 10^{-4}$. For
$\Omega_0$ of this order, the local shape becomes distinguishable
from $\mathbb{R}^{2} \times\mathbb{S}^{1}$ or
$\mathbb{R}\times\mathbb{T}^{2}$. Nevertheless, the bounds shown
here still allow us to severely constrain the magnitude of the
deviation.\cite{next}

Currently one of the most promising methods of searching for
cosmic topology is the so--called circles-in-the-sky method, where
one looks for matching circles of anisotropies in CMB radiation
due to topological identification. For Clifford translations in
flat space these pairs must be strictly antipodal in the celestial
sphere. Hence, in the inflationary limit, the circles due to
isometries of spherical and hyperbolic manifolds should also be
nearly antipolodal. A preliminary search for nearly antipodal
(i.e., with deviation $\theta\leq 10^o$) and with radius larger
than $25^{o}$ was undertaken in Ref. \refcite{Cornish-etal03},
with negative results. It can be shown, however\cite{next}, that
for $|\Omega_0-1|\sim 10^{-4}$ the result (iv) still allows
$\theta\geq10^o$. Also, Ref. \refcite{Roukema} claims to have
found evidence of antipodal circles of radius $<25^{o}$.

\section{Final Remarks}

Given the infinite number of possible candidate manifolds for
cosmic topology, any realistic search strategy must be able to
radically restrict the expected possibilities. This is
particularly important in the inflationary limit, where the order
of any detectable cyclic subgroup as well as the number of
candidate isometries which generate small geodesics are extremely
large.

The results presented here are quite general, imposing severe
constraints on the set of detectable isometries for all detectable
very nearly flat manifolds. By severely restricting the expected
topological signatures of detectable isometries, we are able to
provide an effective framework for searching for evidence of a
non-trivial topology in cosmological observations. More
specifically, these results can be used to confine the parameter
space which realistic search strategies such as the
`circles-in-the-sky' method, need to concentrate on.

\section*{Acknowledgments}

We thank CNPq, CBPF/MCT and FAPERJ for the grants under which this
work was carried out. BM thanks Astronomy unit, QMUL, for
hospitality.

\end{document}